\documentclass[conference]{IEEEtran}

\usepackage{times}
\usepackage{subfigure}
\usepackage{url}
\usepackage{lipsum} 
\usepackage[numbers,comma,sort,sectionbib]{natbib} 

\ifCLASSINFOpdf
   \usepackage[pdftex]{graphicx}
\else
   \usepackage[dvips]{graphicx}
\fi

\usepackage{mathtools}
\DeclarePairedDelimiter\ceil{\lceil}{\rceil}

\begin{document}\sloppy
%
\title{SUIS: An Online Graphical \underline{S}ignature-Based \underline{U}ser \underline{I}dentification \underline{S}ystem}

\author{\IEEEauthorblockN{Shahid Alam}
\IEEEauthorblockA{Department of Computer Science and Engineering\\Qatar University, Doha, Qatar\\
Email: salam@qu.edu.qa}
}


%


\maketitle

\begin{abstract}
Humans possess a large amount of, and almost limitless, visual memory, that assists them to remember pictures far better than words. This phenomenon has recently motivated the computer security researchers' in academia and industry to design and develop graphical user identification systems (GUISs). \emph{Cognometric} GUISs are more memorable than \emph{drawmetric} GUISs, but takes more time to authenticate. None of the previously proposed GUISs combines the advantages of both \emph{cognometric} and \emph{drawmetric} systems. A signature personify a person and a graphical signature is easier to recall than other drawings. This paper proposes a new graphical \underline{S}ignature-based \underline{U}ser \underline{I}dentification \underline{S}ystem named SUIS. It is based on a 2D grid technology, that is used to draw, digitize and store the signature for user identification. SUIS is categorized as both a \emph{cognometric} and \emph{drawmetric} system. Unlike other systems that use 2D grid: We take one cell in a grid as one pixel in the drawing; for signature matching, the signature drawn has to follow the same grid cells as the signature stored, independent of the sequence; and that the system is not based on any machine learning model. Increasing the number of grid cells increases the password space, and decreasing the size of the grid cell increases the precision of the signature. These characteristics makes SUIS: (1) Rigorous enough to be a password system, but easy enough to be usable. (2) Independent of the language and device used to draw the signature. (3) Efficient and practical to be used for online authentication systems.
\end{abstract}


%
\IEEEpeerreviewmaketitle

\section{Introduction and Motivation}\label{sec:introduction}

The traditional methods for user identification systems rely on entering a username and a text password. One of the major vulnerabilities of this technique is the difficulty of remembering passwords. Therefore the users tend to pick short passwords or passwords that are easy to remember. These passwords can be easily broken. Passwords that are hard to break are often hard to remember.

The ability of humans to remember pictures far better than words \cite{picture-memory-1,picture-memory-2} has recently motivated computer security researchers' in academia \cite{gp-survey-acm,gp-survey-ieee,user-authentication-categories,signature-gp,DAS,BDAS,YAGP,web-gp-doodle,pass-go,pass-points,java-gp} and industry \cite{grid-sure,android-pattern-screen-lock,blackberry-pattern-lock,passfaces,commercial-pass-points,windows8-picture-password} to study, design and develop graphical user identification systems (GUISs). More than 10 US patents on GUISs have been issued, the first in 1996 to Greg E. Blonder \cite{gp-blonder} and the last in 2014 to Microsoft Corporation \cite{gp-microsoft-2}, but yet GUISs are not widely used.

There are three basic categories of GUISs \cite{user-authentication-categories}: Cognometrics, Locimetrics, and Drawmetrics. \emph{Cognometric} systems are based on the human cognitive abilities, such as the ability to remember and recall images. \emph{Locimetric} systems are based on locating or identifying a point in an image. \emph{Drawmetric} systems are based on reproducing an already pre-drawn image (outline drawing).

Elizabeth et al. \cite{memory-gp} presents a study about the relationship between memory and graphical passwords. The results indicate that the recognition-based (\emph{cognometric}) graphical passwords are more memorable than recall-based (\emph{drawmetric}) graphical passwords, but takes more time to login. None of the previously proposed GUISs \cite{user-authentication-categories,signature-gp,DAS,BDAS,YAGP,web-gp-doodle,pass-go,pass-points,java-gp,grid-sure,android-pattern-screen-lock,blackberry-pattern-lock,passfaces,commercial-pass-points,windows8-picture-password} combines the advantages of both \emph{cognometric} and \emph{drawmetric} systems. The systems proposed in \cite{signature-gp} and \cite{java-gp} use signature-based schemes but do not use a 2D grid technology (draw, digitize and store the signature), and hence are difficult to recall. The system proposed in \cite{DAS} uses a 2D grid technology but is not a signature-based scheme. Therefore we do not categorize these three \cite{signature-gp,DAS,java-gp} as both \emph{cognometric} and \emph{drawmetric} systems.

Handwritten signatures have long been used as a proof of authorship. In general, signatures are authentic, unforegeable, not reusable, unalterable and unrepudiatable \cite{signature-gp}. Signatures personify a person and are easier to recall than other drawings when drawn on a 2D grid. Therefore we can categorize such signatures as recall-based graphical passwords that have the same or close enough memorability as recognition-based graphical passwords.

In this paper we propose a new graphical \underline{S}ignature-based \underline{U}ser \underline{I}dentification \underline{S}ystem named \textbf{SUIS}. It is based on a 2D grid technology, that is used to draw, digitize and store the signature for user identification. SUIS is categorized as both a \emph{cognometric} and \emph{drawmetric} system. The 2D grid was first used in \emph{Draw a Secret} \cite{DAS} for user identification. Our model is different than proposed in \cite{DAS}:

\begin{itemize}
\item
We take one cell in a grid, as one pixel in the drawing. This makes it much simpler to implement the model and compare the signature in practice.
\item
For signature matching, the signature drawn has to follow the same grid cells as the signature stored, independent of the sequence. This increases the usability of the system, but decreases the password space. Increasing the usability here means, since the users do not have to follow the same sequence they can draw and remember more complex signatures. The password space can be increased by increasing the number of grid cells.
\end{itemize}

Some of the characteristics of SUIS are as follows:

\begin{enumerate}
\item
SUIS is easier and faster to compare for signature matching.
\item
SUIS takes into consideration the angular changes at a coarse level.
\item
SUIS, by using a digitization technique, provides extra security and protection on top of encryption.
\item
The signature window i.e; the grid area (number of grid cells) in SUIS can be increased to increase the password space ($2^n$ where n = number of grid cells) of the system. The size of an average grid (extended) used in our empirical study is $7 \times 7$ for which the password space is $2^{49} > 10 \times 10^{14}$. The password space for a text-based user identification system for $8$ characters (average size) password is $95^{10} > 6 \times 10^{15}$. There are $95$ possible character sets including space.
\item
SUIS is rigorous enough to be a password system, but easy enough to be usable.
\item
Precision of the signature in SUIS can be changed by decreasing the size of the grid cell of the system. For example, a signature drawn with mouse/finger needs less precision, whereas a signature drawn with pen or other such pointing devices needs more precision.
\item
SUIS is independent of the language and the device used to write/draw the signature, but is more suitable for touch-based systems (the empirical study presented in this paper was performed on a touch-based system), such as, mobile devices and laptops.
\item
SUIS is efficient and suitable to be used for online (verification is performed immediately after a password is submitted) authentication systems. Our system is not based on any machine learning model as used in \cite{java-gp}. A machine learning model is not suitable and practical to be used for identifying a user for online authentication systems. Moreover, for successful signature matching such a model needs a large set of training data, i.e; the forged samples of the signatures.
\end{enumerate}

The remainder of the paper is organized as follows. Section \ref{sec:literature-review} describes and compares research efforts that are similar to SUIS. Section \ref{sec:suis} describes SUIS in detail. We conclude in Section \ref{sec:conclusion}.

\section{Related Works}\label{sec:literature-review}

This section discusses some of the previous research efforts on GUISs that are similar to the system proposed in this paper. A thorough survey of these and other systems referenced in Section \ref{sec:introduction} can be found in \cite{gp-survey-acm,gp-survey-ieee}.

Syukri et al. \cite{signature-gp} proposed a system that uses signatures drawn with mouse for identifying a user. Through experimental evaluation they selected parameters for signature verification and matching. User verification threshold was set to 70\%. The size of the signature window was set to $1024 \times 512$ pixels. The distance threshold was set to 50 pixels. The parameters selected for signature verification were:
(1) Number of signature points.
(2) Coordinate of signature points.
(3) Signature writing time.
(4) Signature writing velocity.
(5) Signature writing acceleration.

Experiments were carried out with 21 users. The successful rate achieved was 93\%. To calculate the acceleration they used the classical physics formula $a = \frac{F}{m}$, where $F$ is the user's force to push the mouse and $m$ is the mass of the mouse. Although, the number of participants in the experiments carried out in \cite{signature-gp} were small to make any definite conclusion, the successful rate reported is very encouraging, and indicates that some of the parameters can be used for a successful signature matching.

SUIS indirectly uses the signature points as part of the grid cells. We do not use the last three parameters selected by \cite{signature-gp}, as we believe these are not the true representation of a user drawing a signature. In practice the last three parameters are more dependent than the first two parameters, on different environments, such as the mood of the user (e.g; in sickness, sadness and excitement etc), times of the day or night, etc, and hence can produce more true negatives.

\begin{figure*}[htbp]
\centering
   {\includegraphics[scale=0.65]{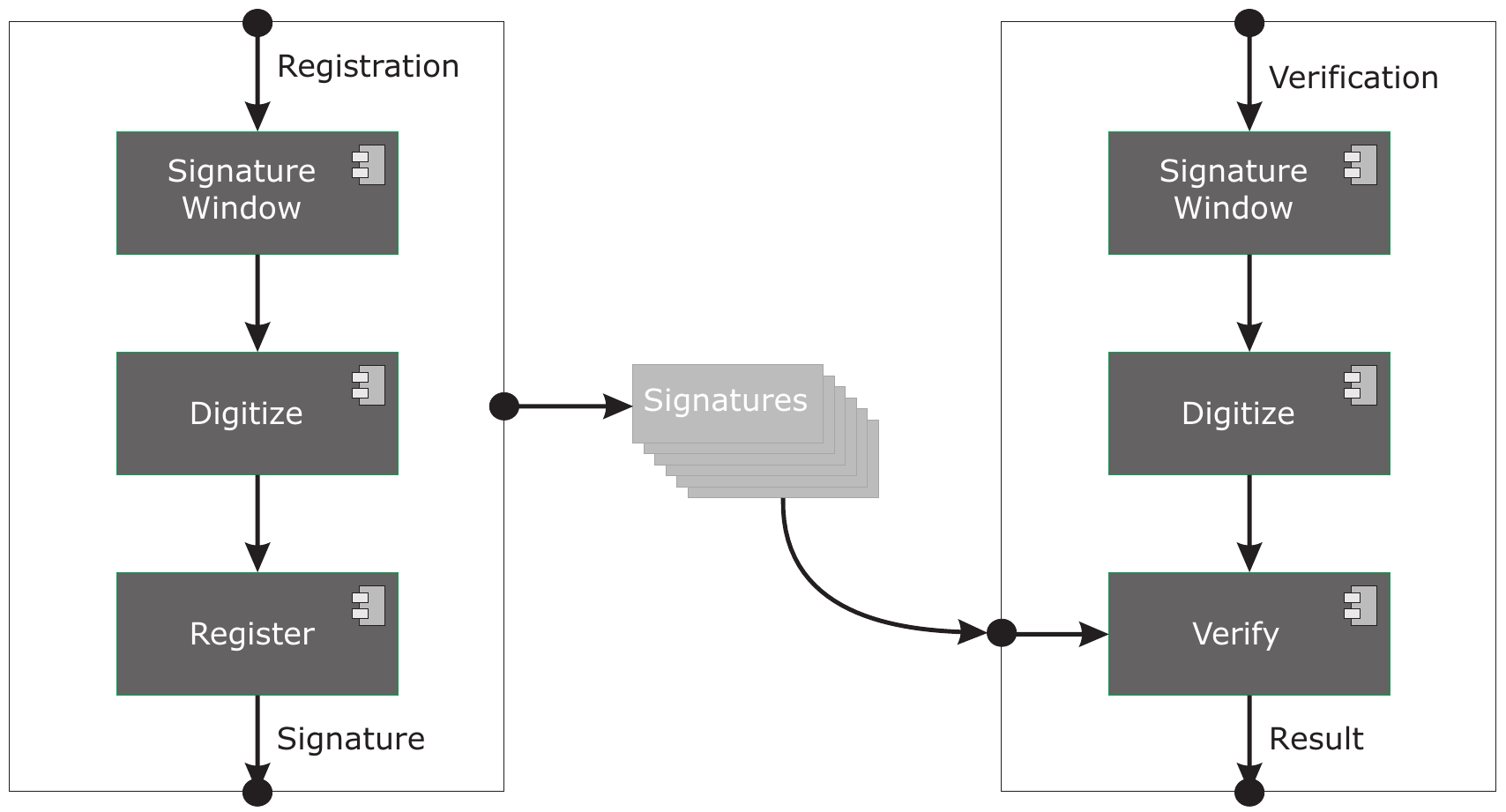}}
\caption{High level overview of SUIS (Signature-Based User Identification System).}
\label{fig:suis-system-overview}
\end{figure*}

Jermyn et al. \cite{DAS} proposed a system called \emph{Draw a Secret} (DAS), that allows the user to draw a unique password. The password is drawn on a 2D grid. If the stored and the drawn password touches the same grid cells in the same sequence, then the user is authenticated. It becomes difficult to authenticate if the strokes of the user are too close to the grid lines. In this case either, the user is presented with the internal representation to confirm if the cells were actually touched by the drawing, or, the system does not accept a drawing that is too close to a grid line.

As passwords depend on users, the proposed system in \cite{DAS} lacks an empirical study for its usability. It has only been tested through paper prototypes. Because of lack of a suitable user study we cannot comment on its effectiveness as a GUIS.

Everitt et al. \cite{java-gp} proposed a neural network-based system using graphical signatures. The system use a hybrid approach, using both text and graphical passwords. The input devices used are mouse for graphical-based password and keyboard for text-based password. The authenticity of a user is confirmed by the typing style and the signature match. For the typing style they measure two times, one is the time between the two key presses, and the other is the time a key is held down. This idea for the key metrics is similar in concept to the one used in \cite{signature-gp} for the mouse metrics. For matching two graphical signatures they use the signature traces, and measure the change in angles and euclidean distances in the two signatures.

The experiments were carried out with 41 participants between the ages of 20 and 30. The results show that they achieved a false accept rate (FAR) of 4.4\% and a range of false reject rate (FRR) from 0.2\% -- 38.6\%. FAR is the rate at which forged samples are accepted as genuine and FRR is the rate at which genuine samples are rejected as forgeries.

The system \cite{java-gp} is based on a machine learning model and hence needs training in addition to registration and verification. The training data is usually created by asking users to provide a set of forged signature samples for other users or these forge signature samples are generated automatically. We think this is one of the major problems of using a machine learning model in such systems.

\section{SUIS (Signature-Based User Identification System)}\label{sec:suis}

Figure \ref{fig:suis-system-overview} gives an overview of SUIS proposed in this paper. SUIS has two phases. In the first phase the user registers her/his signature. This signature, after digitizing\footnote{Digitizing also includes encryption. We do not discuss it in this paper because there is already a lot of literature available on encryption.}, is stored in the database. In the second phase the signature drawn by the user, after digitizing, is verified against the stored signatures. In case of a match the user is identified as a legal user.

\subsection{The Digitization Technique}\label{sec:digitization}

We use a simple but practical digitization technique to store the signature. This makes the signature easy to store and compare, but difficult enough to make its extraction non-trivial.

In SUIS, size (number of grid cells) of the 2D grid for drawing the signature range from 25 (5 x 5) -- 49 (7 x 7), to keep the password space in the range of more than ten million ($2^{25} > 1 \times 10^{7}$) to more than a trillion ($2^{49} > 1 \times 10^{14}$). Each cell in the 2D grid is stored as part of the signature. There are two grids: the \emph{drawing grid}, that is visible to the user for drawing, and the \emph{extended grid} that includes the \emph{drawing grid} and extra cells, and is used to digitize and store the signature. If a cell in a \emph{drawing grid} contains a drawing, i.e, the pixels touched in the cell are inside the drawing area of the cell, it is stored as 1, otherwise it is stored as 0. To produce a coarser signature, we keep the drawing area in a cell smaller than the area of the cell. The extra cells in the \emph{extended grid} is used to store the value of the color selected by the user and the value of the randomly selected storing technique used. The ability of selecting a color and choosing a random value for the storing technique for the signature also increase the password space.

The user can select a color to draw the signature. To elude shoulder surfing upto an extent, a different value of the color relative to the color selected (in which the signature drawn is displayed) is selected and stored. Shoulder surfing is a technique where a person looks over someone's shoulder to get information, such as passwords, PINs, other security codes and data.  Each color in SUIS is assigned a number. The value of color stored is computed as follows: $color \ stored = number \ assigned \ to \ the \ selected \ color + \ceil*{\frac{N}{t}}$, where $N$ is the number of total colors used in the system, $t$ is the randomly selected storing technique used and $N$ is $\geq$ $t$. To provide more resistance for shoulder surfing, we erase the signature as soon as it's drawn completely and the user submits it for verification.

We use a number of different techniques for storing a signature. To make the extraction of the information about a signature (grid cells drawn, value of color and storing technique) non-trivial, each time a signature is stored a different storing technique is selected randomly. This information is stored as part of the signature, and also as part of the user's profile, so that during the verification phase of the user the same storing technique is used to verify the signature. Each time, when a user login, a different technique can be used and stored in the user's profile to increase the protection of the user's signature. We also encrypt the signature before storing it. Therefore, this digitization of the signature provides extra security and protection on top of encryption.

\subsection{Signature Storing Techniques}\label{sec:storing-techniques}


A signature is stored as a 2D matrix. Each cell in the matrix contains either a value 1 or a value 0. Initially all the cells in the matrix contains a value 0. The signature drawn by the user is stored in the cells of the matrix corresponding to the \emph{drawing grid}, as explained above. The \emph{extended grid} contains twice (to take care of $\ceil*{\frac{N}{t}}$) as many extra cells as the total number of colors that can be used to draw the signature. Based on the value of the color selected, the corresponding cell out of these cells of the matrix gets the value of 1. Similarly the \emph{extended grid} contains another set of additional cells equal to the number of the storing techniques available. Based on the storing technique used, the corresponding cell out of these additional cells of the matrix gets the value of 1.

Each signature's storing technique is given a number (a value) that is stored in the signature, as explained above. We introduce \emph{simple -- complex} changes in each signature technique to make it different than the other. Some of the changes introduced are:

\begin{enumerate}
\item
Changing the numbering of the signature's matrix (from left-right to right-left and so on).
\item
Changing the location (start, middle or the end) where the value of the color is stored in the signature's matrix.
\item
Changing the location (start, middle or the end) where the value of the storing technique used is stored in the signature's matrix.
\item
Instead of storing a 1 with 0's, we could just store the value of the color and the value of the storing technique in the matrix.
\item
By splitting or merging the grid cells and storing them at different locations in the signature's matrix.
\item
By just storing the information about either 0's, 1's or both in the signature's matrix. For example storing only the location of all the 1's in the matrix.
\item
Combination of two or more of the above techniques.
\end{enumerate}

Other complex storing techniques can also be used to increase protection, such as matrix manipulations, etc, and we leave this to the reader. Using different number of storing techniques gives SUIS the ability to randomize the value of the storing technique, each time a signature is stored, that makes it non-trivial to extract the signature information.

\subsection{Example}\label{sec:example}

\begin{figure}[htbp]
\centering
   {\includegraphics[scale=0.45]{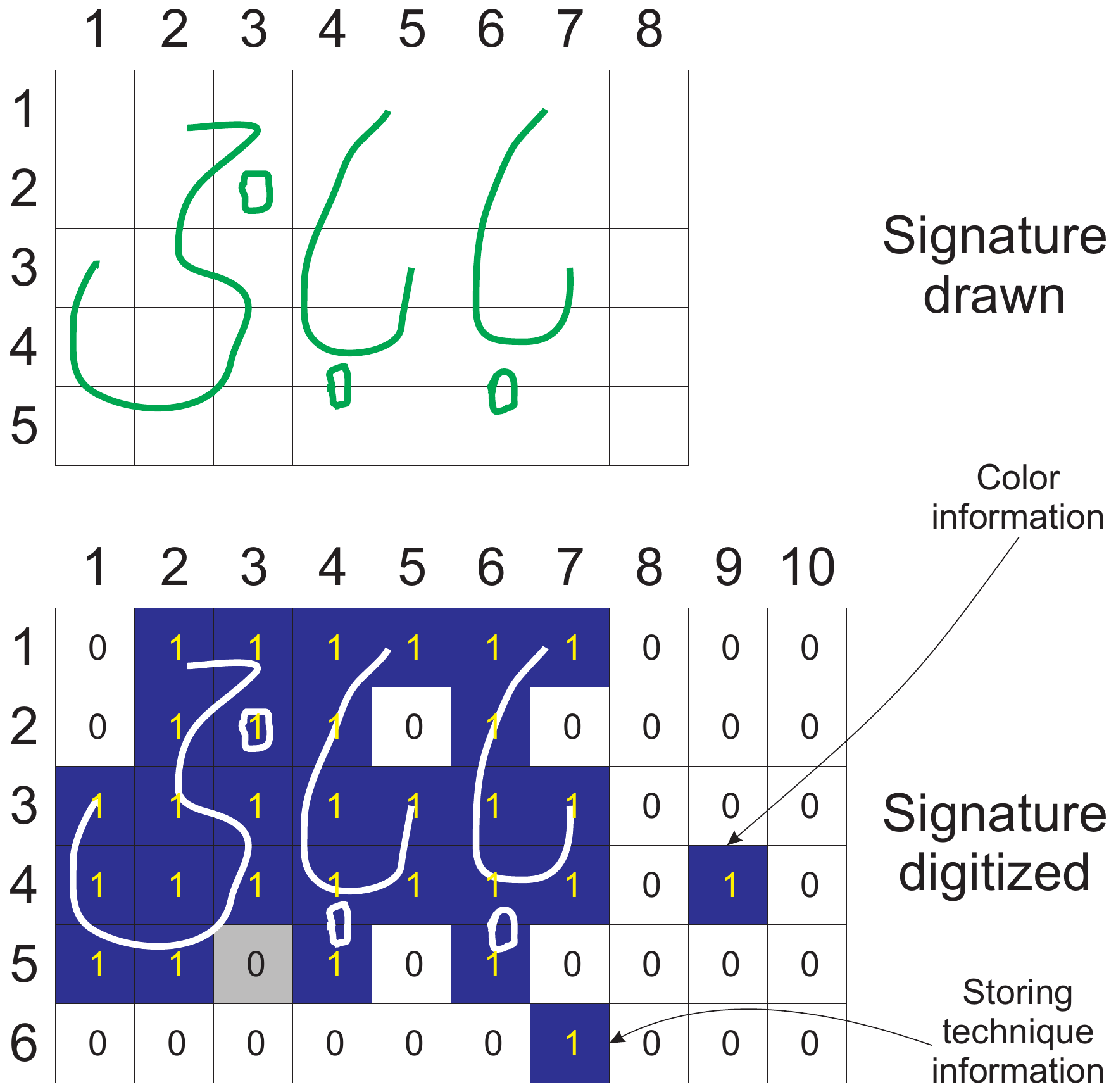}}
\caption{An example of digitization technique used in SUIS. The signature \emph{Babajee} (as pronounced in English language), a word used in Urdu and Hindi languages to \emph{refer an old man with respect}, is digitized.}
\label{fig:suis-digitization}
\end{figure}

We explain the digitization technique described in Section \ref{sec:digitization} and how it is used for storing (using one of the randomly selected storing techniques) and matching a signature, using an example shown in Figure \ref{fig:suis-digitization}.

A 2D grid of $8 \times 5$ is used in Figure \ref{fig:suis-digitization} for drawing, called the \emph{drawing grid}. The same 2D grid is extended to $10 \times 6$ to store the signature, called the \emph{extended grid}. Whenever the signature touches a grid cell, a value of 1 is stored in that grid cell. We store a value of 0 in the grid cell at column 3 and row 5 (shown in \emph{lightgray} color), because the number of pixels touched in the grid cell are less than a predefined threshold value for this grid.

The user, used color \emph{green} to draw the signature and the storing technique (randomly selected) used for storing the signature is numbered 1. So we store number 7 (assuming color \emph{green} = 1 and color \emph{white} = 7, 1 + $\ceil*{\frac{16}{3}} = 7$) and 1 in the \emph{extended grid}, at the end of the matrix (the last 2 columns and the last row). For storing the color number 7, we store a 1 at the corresponding cell, at column 9 and row 4, in the matrix. For this example we have assumed there are 4 signature storing techniques available in SUIS, so the \emph{extended grid} has 16 + 4 = 20 extra cells. For storing the signature's storing technique number 1, we store a 1 at the corresponding cell, at column 7 and row 6, in the matrix, and the same number is also stored as part of the user's profile.

For matching a signature, we exactly match all the extra cells added to the \emph{extended grid} of all the stored signatures with all the extra cells added to the \emph{extended grid} of the signature drawn for the verification. In case of a successful match we match the corresponding \emph{drawing grid} of the stored signature with the \emph{drawing grid} of the signature drawn for the verification, based on a predefined threshold value for the grid. If the difference is $\geq$ to the predefined threshold value the match is successful, and the user is verified as a legal user.





\section{Conclusion and Future Work}\label{sec:conclusion}

In this paper, we have proposed a new graphical signature-based user identification system, that is efficient and practical to be used in login systems. The system is rigorous enough to be a password, but easy enough to be usable. It is independent of the language used to write/draw the signature. 

Currently we are developing a prototype tool in Java to implement SUIS. In future we will carry out a study with a large number of participants and evaluate the validity of SUIS using a number of metrics, such as \emph{Usability}, \emph{Deployability} and \emph{Security} described in \cite{Evaluation-Web-Authentication}. To improve the usability of the current technique, in future, we will develop different techniques, such as by integrating joining blocks/lines that will make it easier to form a shape or a drawing as a graphical signature, etc, and test the pros and cons of each such technique through a large scale empirical study.



\end{document}